\documentclass[aps,prl,superscriptaddress,showpacs,twocolumn,floatfix]{revtex4}

\usepackage{graphicx}

\def\deg{^\circ}
\def\gtorder{\mathrel{\raise.3ex\hbox{$>$}\mkern-14mu
 \lower0.6ex\hbox{$\sim$}}}
\def\ltorder{\mathrel{\raise.3ex\hbox{$<$}\mkern-14mu
 \lower0.6ex\hbox{$\sim$}}}
\def\f20{F_2^{(0)}}

\begin{document}

\title{Scaling of the $F_2$ structure function in nuclei and quark distributions at $x>1$}

\author{N.~Fomin}
\affiliation{University of Virginia, Charlottesville, VA, USA}
\affiliation{University of Tennessee, Knoxville, TN, USA}

\author{J.~Arrington}
\affiliation{Physics Division, Argonne National Laboratory, Argonne, IL, USA}

\author{D.~B.~Day}
\affiliation{University of Virginia, Charlottesville, VA, USA}

\author{D.~Gaskell}
\affiliation{Thomas Jefferson National Laboratory, Newport News, VA, USA}

\author{A.~Daniel}
\affiliation{University of Houston, Houston, TX, USA}

\author{J.~Seely}
\affiliation{Massachusetts Institute of Technology, Cambridge, MA, USA}

\author{R.~Asaturyan}
\thanks{Deceased}
\affiliation{Yerevan Physics Institute, Armenia}

\author{F.~Benmokhtar}
\affiliation{University of Maryland, College Park, MD, USA}

\author{W.~Boeglin}
\affiliation{Florida International University, Miami, FL, USA}

\author{B.~Boillat}
\affiliation{Basel University, Basel, Switzerland}

\author{P.~Bosted}
\affiliation{Thomas Jefferson National Laboratory, Newport News, VA, USA}

\author{A.~Bruell}
\affiliation{Thomas Jefferson National Laboratory, Newport News, VA, USA}

\author{M.~H.~S.~Bukhari}
\affiliation{University of Houston, Houston, TX, USA}

\author{M.~E.~Christy}
\affiliation{Hampton University, Hampton, VA, USA}

\author{E.~Chudakov}
\affiliation{Thomas Jefferson National Laboratory, Newport News, VA, USA}

\author{B.~Clasie}
\affiliation{Massachusetts Institute of Technology, Cambridge, MA, USA}

\author{S.~H.~Connell}
\affiliation{University of Johannesburg, South Africa}

\author{M.~M.~Dalton}
\affiliation{University of Virginia, Charlottesville, VA, USA}

\author{D.~Dutta}
\affiliation{Mississippi State University, Jackson, MS, USA}
\affiliation{Duke University, Durham, NC, USA}

\author{R.~Ent}
\affiliation{Thomas Jefferson National Laboratory, Newport News, VA, USA}

\author{L.~El Fassi}
\affiliation{Physics Division, Argonne National Laboratory, Argonne, IL, USA}

\author{H.~Fenker}
\affiliation{Thomas Jefferson National Laboratory, Newport News, VA, USA}

\author{B.~W.~Filippone}
\affiliation{Kellogg Radiation Laboratory, California Institute of Technology, Pasadena, CA, USA}

\author{K.~Garrow}
\affiliation{TRIUMF, Vancouver, British Columbia, Canada}

\author{C.~Hill}
\affiliation{University of Virginia, Charlottesville, VA, USA}

\author{R.~J.~Holt}
\affiliation{Physics Division, Argonne National Laboratory, Argonne, IL, USA}

\author{T.~Horn}
\affiliation{University of Maryland, College Park, MD, USA}
\affiliation{Thomas Jefferson National Laboratory, Newport News, VA, USA}

\author{M.~K.~Jones}
\affiliation{Thomas Jefferson National Laboratory, Newport News, VA, USA}

\author{J.~Jourdan}
\affiliation{Basel University, Basel, Switzerland}

\author{N.~Kalantarians}
\affiliation{University of Houston, Houston, TX, USA}

\author{C.~E.~Keppel}
\affiliation{Thomas Jefferson National Laboratory, Newport News, VA, USA}
\affiliation{Hampton University, Hampton, VA, USA}

\author{D.~Kiselev}
\affiliation{Basel University, Basel, Switzerland}

\author{M.~Kotulla}
\affiliation{Basel University, Basel, Switzerland}

\author{R.~Lindgren}
\affiliation{University of Virginia, Charlottesville, VA, USA}

\author{A.~F.~Lung}
\affiliation{Thomas Jefferson National Laboratory, Newport News, VA, USA}

\author{S.~Malace}
\affiliation{Hampton University, Hampton, VA, USA}

\author{P.~Markowitz}
\affiliation{Florida International University, Miami, FL, USA}

\author{P.~McKee}
\affiliation{University of Virginia, Charlottesville, VA, USA}

\author{D.~G.~Meekins}
\affiliation{Thomas Jefferson National Laboratory, Newport News, VA, USA}

\author{T.~Miyoshi}
\affiliation{Tohoku University, Sendai, Japan}

\author{H.~Mkrtchyan}
\affiliation{Yerevan Physics Institute, Armenia}

\author{T.~Navasardyan}
\affiliation{Yerevan Physics Institute, Armenia}

\author{G.~Niculescu}
\affiliation{James Madison University, Harrisonburg, VA, USA}

\author{Y.~Okayasu}
\affiliation{Tohoku University, Sendai, Japan}

\author{A.~K.~Opper}
\affiliation{Ohio University, Athens, OH, USA}

\author{C.~Perdrisat}
\affiliation{College of William and Mary, Williamsburg, VA, USA}

\author{D.~H.~Potterveld}
\affiliation{Physics Division, Argonne National Laboratory, Argonne, IL, USA}

\author{V.~Punjabi}
\affiliation{Norfolk State University, Norfolk, VA, USA}

\author{X.~Qian}
\affiliation{Duke University, Durham, NC, USA}

\author{P.~E.~Reimer}
\affiliation{Physics Division, Argonne National Laboratory, Argonne, IL, USA}

\author{J.~Roche}
\affiliation{Ohio University, Athens, OH, USA}
\affiliation{Thomas Jefferson National Laboratory, Newport News, VA, USA}

\author{V.M.~Rodriguez}
\affiliation{University of Houston, Houston, TX, USA}

\author{O.~Rondon}
\affiliation{University of Virginia, Charlottesville, VA, USA}

\author{E.~Schulte}
\affiliation{Physics Division, Argonne National Laboratory, Argonne, IL, USA}

\author{E.~Segbefia}
\affiliation{Hampton University, Hampton, VA, USA}

\author{K.~Slifer}
\affiliation{University of Virginia, Charlottesville, VA, USA}

\author{G.~R.~Smith}
\affiliation{Thomas Jefferson National Laboratory, Newport News, VA, USA}

\author{P.~Solvignon}
\affiliation{Physics Division, Argonne National Laboratory, Argonne, IL, USA}

\author{V.~Tadevosyan}
\affiliation{Yerevan Physics Institute, Armenia}

\author{S.~Tajima}
\affiliation{University of Virginia, Charlottesville, VA, USA}

\author{L.~Tang}
\affiliation{Thomas Jefferson National Laboratory, Newport News, VA, USA}
\affiliation{Hampton University, Hampton, VA, USA}

\author{G.~Testa}
\affiliation{Basel University, Basel, Switzerland}

\author{R.~Trojer}
\affiliation{Basel University, Basel, Switzerland}

\author{V.~Tvaskis}
\affiliation{Hampton University, Hampton, VA, USA}

\author{W.~F.~Vulcan}
\affiliation{Thomas Jefferson National Laboratory, Newport News, VA, USA}

\author{C.~Wasko}
\affiliation{University of Virginia, Charlottesville, VA, USA}

\author{F.~R.~Wesselmann}
\affiliation{Norfolk State University, Norfolk, VA, USA}

\author{S.~A.~Wood}
\affiliation{Thomas Jefferson National Laboratory, Newport News, VA, USA}

\author{J.~Wright}
\affiliation{University of Virginia, Charlottesville, VA, USA}

\author{X.~Zheng}
\affiliation{University of Virginia, Charlottesville, VA, USA}
\affiliation{Physics Division, Argonne National Laboratory, Argonne, IL, USA}

\date{\today}

\begin{abstract}

We present new data on electron scattering from a range of nuclei taken
in Hall C at Jefferson Lab.  For heavy nuclei, we observe a rapid falloff in
the cross section for $x>1$, which is sensitive to short range contributions
to the nuclear wave-function, and in deep inelastic scattering corresponds to
probing extremely high momentum quarks.  This result agrees with higher energy
muon scattering measurements, but is in sharp contrast to neutrino scattering
measurements which suggested a dramatic enhancement in the distribution of the
`super-fast' quarks probed at $x>1$.  The falloff at $x>1$ is noticeably
stronger in $^2$H and $^3$He, but nearly identical for all heavier nuclei.

\end{abstract}



\maketitle


The quark structure of nuclei is extremely complex, and a detailed understanding
of nuclei at the quark level requires the careful incorporation of nucleonic
degrees of freedom and interactions as well as the dynamics of quarks and
gluons.  In inclusive electron scattering from nuclei, the cross section is
characterized by the structure functions $F_1(\nu,Q^2)$ and $F_2(\nu,Q^2)$,
where $\nu$ is the energy transfer and $-Q^2$ is the square of the
four-momentum transfer.  At high $Q^2$, the reaction is dominated by elastic
scattering from quasi-free quarks, and one can probe the momentum distribution
of the quarks.  In the Bjorken limit ($\nu$, $Q^2 \to \infty$), the quark mass
and transverse momenta are negligible compared to the energy and momentum of
the probe, and the scattering is sensitive only to the the quark longitudinal
momentum, where $x=Q^2/(2M\nu)$ is the fraction of the hadrons longitudinal
momentum carried by the quark in the infinite momentum frame. In this deep
inelastic scattering (DIS) limit, the structure functions exhibit
\textit{scaling}, i.e. $F_2(\nu,Q^2) \to F_2(x)$, becoming independent of
$Q^2$ at fixed $x$, with $F_2(x)$ proportional to a charge-weighted sum of the
quark momentum distributions in the target.

As one moves away from the Bjorken limit, there are deviations from perfect
scaling.  At finite-$Q^2$, kinematical corrections yield a $Q^2$ dependence
that can be large for low $Q^2$ or large $x$.  While these scaling
violations have historically been called ``target mass''
corrections~\cite{schienbein08}, they are in fact independent of the mass of
the target for a quark of a given longitudinal momentum. At lower $Q^2$, there
are also significant contributions from higher twist effects, e.g. structure
due to quark--quark and quark--gluon correlations which appear most clearly as
strong resonance structure.  These scaling violating terms make extraction of
the quark distributions most straightforward at high energies. QCD evolution
yields an approximately logarithmic $Q^2$ dependence at all $Q^2$ values, but
this is a true scale dependence of the parton distributions.

The early expectation was that the nuclear quark momentum distribution would
be a convolution of the distribution of nucleons in a nucleus with the the
distribution of quarks in the nucleons.  Contrary to these expectations,
measurements of inclusive scattering from nuclei showed a 10--20\%
suppression of high momentum quarks ($0.3<x<0.8$) in heavy
nuclei~\cite{geesaman95}, demonstrating that the quark distributions in nuclei
are not simply a sum of the proton and neutron's quark distributions.

The quark distribution at $x>1$ is extremely small in the convolution
model, as the nucleon quark distributions fall rapidly as $x \to 1$ and there
are very few fast nucleons available to boost the quarks to $x > 1$.
The bulk of these `super-fast' quarks come from nucleons above the
Fermi momentum, which are generated by the strongly repulsive core of the N--N
interaction; they reflect the short-range correlations (SRCs) in the ground
state nuclear wave-function~\cite{frankfurt93, sargsian03}.  Exotic
configurations, such as 6-quark bags, may provide an even more efficient
mechanism for generating very high momentum quarks, as the quarks from two
nucleons can more freely share momentum~\cite{geesaman95, sargsian03}.  It is
clear that a holistic explanation of DIS from nuclei must describe the
behavior of the structure functions in the full kinematic range, and
measurements at $x>1$ are necessary to illuminate the presence of short range
structure in nuclei.

There are only two high energy measurements thus far for $x \gtorder 1$, and
they yield dramatically different results.  Muon scattering data from the BCDMS
collaboration~\cite{benvenuti94} for $Q^2$ from 52--200~GeV$^2$ show a rapid
falloff in $F_2(x)$.  They find $F_2(x) \propto \exp(-s \cdot x)$ with $s=16.5
\pm 0.6$ for $0.75<x<1.05$, which suggests the need for relatively modest
contributions from short-range correlations.  Neutrino scattering measurements
from the CCFR collaboration~\cite{vakili00} at $Q^2=125$~GeV$^2$ yield
$s=8.3 \pm 0.7$ for $0.75<x<1.2$, which has been interpreted as an
indication of exceptionally large strength from short-range correlations or
the need for other, more exotic, contributions. However, both measurements
have important limitations: CCFR had to make significant corrections due to
the poor resolution on their reconstructed $x$ value, while BCDMS was only
able to extract $F_2$ up to $x=1.05$.  It is unclear if this is sufficient
to make meaningful comparisons to model predictions of short range structure
in nuclei, as this is not expected to dominate until $x \gtorder
1.2$~\cite{rozynek88, frankfurt88}. More extensive measurements have been made
at lower energies, but they have been limited to $x \approx 1$~\cite{bosted92,
arrington96} or $Q^2 < 5$~GeV$^2$~\cite{day79, filippone92, arrington01}.

We present the results of JLab E02-019, which made new measurements of
inclusive scattering from nuclei, covering an expanded range in $x$ and $Q^2$.
Data were taken for few-body and heavy nuclei, covering both the region of the
EMC effect~\cite{seely09} and $x>1$. The measurement was performed in Hall C
at the Thomas Jefferson National Accelerator Facility in 2004.  A continuous
wave electron beam of 5.766~GeV and current of $\approx$80 $\mu$A was
supplied. Electrons scattered from the target were detected using the High
Momentum Spectrometer (HMS) at $\theta =$~18$^{\circ}$, 22$^{\circ}$,
26$^{\circ}$, 32$^{\circ}$, 40$^{\circ}$, and 50$^{\circ}$, corresponding to
$2 \ltorder Q^2 \ltorder 9$~GeV$^2$.  Data were taken on cryogenic $^2$H,
$^3$He, $^4$He targets, solid Be, C, Cu, and Au targets, as well as Al targets
used to measure and correct for the contribution from the walls of the
cryogenic target cells.

Electrons were selected using the HMS gas \v{C}erenkov and electromagnetic
calorimeter detectors with efficiencies of $>$98\% and $>$99.7\%, respectively,
and negligible pion contamination.  Data were also taken with the HMS in
positive polarity to determine the contribution from
charge-symmetric processes.  The systematic error associated with positron
subtraction was negligible except at the largest $Q^2$ values, where it is
under 2\%.  The uncertainty due to the spectrometer acceptance is 1.4\%.

The largest sources of systematic uncertainty at high $x$ come from
the beam energy (0.05\%), HMS central momentum (0.05\%) and angle
(0.5mr) settings. The impact of these on the cross
section is typically small (1--2\% for $x<1$) for all but the largest $x$
values, where the uncertainty can reach 4--6\%. The
systematic uncertainty that arises from subtraction of the aluminum end-caps
(cryogenic targets only) also grows with increasing $x$ values, as does the
relative contribution of the aluminum end-caps to the measured cross
section, giving a range in error of 0.3-2.4\%. The cross sections also had to be corrected for
bin-centering, radiative, and Coulomb effects.  The systematic errors
associated with those corrections are 0.5\%, 1.4\%, and $<$2\%, respectively. 
Details on the analysis and uncertainties, as well as the various corrections
applied can be found in Ref.~\cite{fominphd}. The total systematic uncertainty
on the extracted cross sections is below 4\% for $x<1$, and up to 6\% in the
$1< x \lesssim 2$ range.

The structure function per nucleon, $F_2(\nu,Q^2)$, is extracted from the
measured cross section as follows:
\begin{equation}
F_2=\frac{d^2\sigma}{d\Omega dE}\cdot \frac{\nu}
{\sigma_{mott}[1+2\: \tan^2(\theta /2)\frac{1+\nu^2/Q^2}{1+R}]}\:\rm{,}
\end{equation}
with $R$=$\sigma_L/\sigma_T$=$(0.32$~GeV$^2)$/$Q^2$~\cite{bosted92} and
$\delta R/R$=50\%, yielding an additional uncertainty of $\ltorder$1\% in $F_2$.

\begin{figure}[htpb] 
\includegraphics[height=0.44\textwidth,width=4.4cm,angle=270]{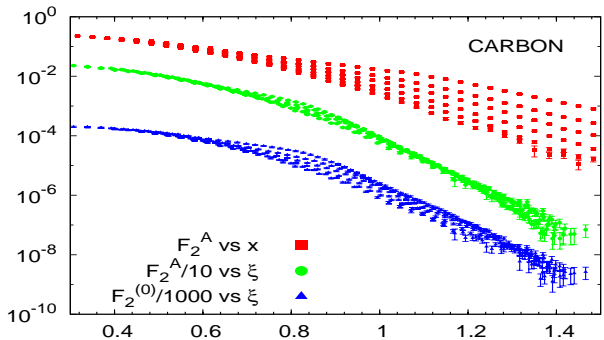}
\caption{(Color online) $F_2$ for the E02-019 carbon data ($2 \ltorder Q^2
\ltorder 9$~GeV$^2$) as a function of $x$ (top) and $\xi$ (middle), and $\f20$
vs $\xi$ (bottom).  In each case, the higher points correspond to the smaller
scattering angles (lowest $Q^2$ values).}
\label{fig:all_f2s}
\end{figure}

Scaling of the proton structure functions has been observed over a large
kinematic range in high energy inclusive scattering.  Based on these
measurements, the DIS region is often taken to be $W^2>4$~GeV$^2$,
$Q^2>1$~GeV$^2$, where $W^2 = M_p^2 + 2M_p\nu - Q^2$ is the square of the
invariant mass of the undetected hadronic system and $M_p$ is the proton mass.
It has been shown that scaling violations are reduced when one examines $F_2$
at fixed $\xi={2x}/(1+r)$~\cite{nachtmann73}, where $r=\sqrt{1+Q^2/\nu^2}$. 
$\xi$ is equivalent to $x$ in the Bjorken limit, but when examining
scaling at fixed $\xi$, rather than fixed $x$, the observed scaling behavior
extends to lower $W^2$~\cite{filippone92, arrington01}, corresponding to
larger $\xi$ values.  This improved scaling can be seen clearly in
Fig.~\ref{fig:all_f2s}, where the upper sets of curves show $F_2$ for carbon
plotted against $x$ (red squares) and $\xi$ (green circles) over a range of
$Q^2$ values.  The extended $\xi$-scaling of the nuclear structure function,
seen to begin above 3--4~GeV$^2$, may allow access to quark distributions for
$\xi \gtorder 1$~\cite{arrington06a, melnitchouk05}.

We compare our data to higher $Q^2$ measurements, using a partonic framework
to look for deviations from the scaling picture. Rather than simply examining
$F_2$ as a function of $\xi$, as was done previously, we account
for the kinematical scaling violations using the prescription of
Ref.~\cite{schienbein08} (Eq.(23)) and study the scaling of $\f20(\xi,Q^2$):
\begin{eqnarray}
\label{eq:tmc}
\frac{x^2}{\xi^2r^3}\f20(\xi,Q^2) & =
& F_2(x,Q^2) - \frac{6M^2x^3}{Q^2r^4}h_2(\xi,Q^2)
\nonumber \\ 
&& - \frac{12M^4x^4}{Q^4r^5}g_2(\xi,Q^2),
\end{eqnarray}
where $h_2(\xi,Q^2)\!=\!\int_{\xi}^A u^{-2} \f20(u,Q^2) du$ and
$g_2(\xi,Q^2)\!=\!\int_{\xi}^A v^{-2} (v-\xi) \f20(v,Q^2) dv$.  One could
incorporate these effects into a partonic model for $F_2$, rather than
extracting an ``idealized'' scaling function, but this approach
minimizes the $Q^2$ dependence in the scaling function, making it easier to
directly compare different data sets.

\begin{figure}[!htpb]
\begin{center}
\includegraphics[height=0.47\textwidth,width=9cm,angle=270]{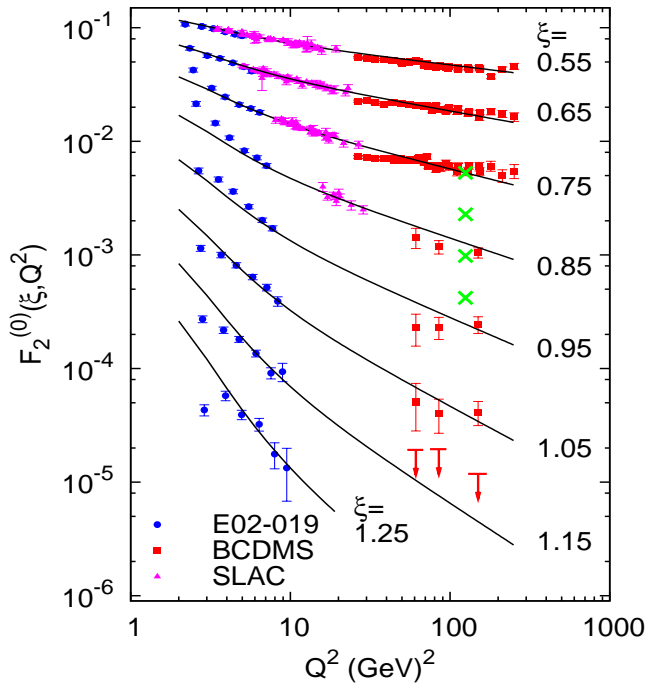}
\caption{(Color online) $\f20$ vs $Q^2$ for fixed $\xi$ value.  For this work
and BCDMS, the carbon data are shown, while the SLAC points are carbon
pseudo-data taken from measurements on deuterium.  The solid
curves are the global fit, the short horizontal red lines show the BCDMS
$\xi$=1.15 upper limit, and the green crosses show the falloff between
$\xi$=0.75 and $\xi$=1.05 based on the CCFR data (see text for details).}
\label{fig:f20global}
\end{center}
\end{figure}

To calculate $h_2$ and $g_2$, we use a factorized model for $\f20(\xi,Q^2)$,
with a common $Q^2$ dependence for all targets and a simple fit to
$\f20(\xi,Q_0^2)$ for each nucleus. In the partonic picture, the $Q^2$
dependence should come only from QCD evolution.  Because we cannot determine
QCD evolution without knowing the partonic structure, we fit the the 
$Q^2$ dependence of the world's data (shown in Fig.~\ref{fig:f20global}),
excluding our lower $Q^2$ points, at several $\xi$ values to a functional form
chosen to be consistent with evolution.  This is used to scale our new
data to $Q_0^2=7$~GeV$^2$, and we obtain a fit for $\f20(\xi,Q_0^2)$ from a
subset of these data, chosen to minimize contributions from quasielastic
scattering. This simple fit provides a reasonable description of the global
data set (see Fig.~\ref{fig:f20global}), with deviations at low $Q^2$, in
particular near the quasielastic peak ($\xi \approx 0.85$) and for the largest
values of $\xi$. The $h_2$ and $g_2$ terms yield corrections of up to 15\% at
the lower $Q^2$ values, but $\ltorder$5\% for $Q^2>5$~GeV$^2$.  We estimate
the model dependence in the extraction of $\f20$ to be $\ltorder$2\%.

Figure~\ref{fig:all_f2s} shows $\f20$ vs $\xi$ (blue triangles), which
has somewhat \textit{greater} $Q^2$ dependence at large $\xi$ than the
uncorrected structure function $F_2$ (green circles).  The main difference
between $\xi$-scaling and our scaling analysis is the factor
$x^2/(\xi^2 r^3)$ in front of the leading term in Eq.~\ref{eq:tmc}, as the
$h_2$ and $g_2$ terms are relatively small.  Neglecting this pre-factor 
introduces an additional $Q^2$ dependence that approximately cancels that of
the QCD evolution, resulting in an artificially small $Q^2$ dependence in
the naive $\xi$-scaling analysis.

Figure~\ref{fig:f20global} shows the carbon results for $\f20(\xi,Q^2)$,
scaled to fixed values of $\xi$ using our global fit.  The SLAC points are
deuterium data~\cite{whitlow92}, multiplied by the SLAC E139~\cite{gomez94}
fit to the carbon-to-deuteron structure function ratio, yielding carbon
pseudo-data to provide a continuous $Q^2$ range for lower $\xi$ values.  For
all data sets, $\f20$ is extracted from the measured structure functions using
the global fit to calculate $g_2$ and $h_2$. For $\xi \le 0.75$, where the
high $Q^2$ data determine the evolution, our data are in excellent agreement
with this $Q^2$ dependence down to $Q^2=3$~GeV$^2$.  The observed $Q^2$
dependence grows slowly with $\xi$ over this region, and with a continued
increase at higher $\xi$ values, our highest $Q^2$ measurements are consistent
with SLAC and BCDMS. For large $\xi$ values at low $Q^2$, our data deviate
from this $Q^2$ dependence due to higher twist contributions, especially in
the vicinity of the quasielastic peak ($\xi \approx 0.85$)

The CCFR measurement did not explicitly extract values of $F_2$, but obtained
a fit to the falloff at large $\xi$.  We illustrate this falloff by
normalizing to our global fit at $\xi=0.75$ and applying the CCFR $\xi$
dependence to extract $\f20$ at $\xi=0.75$, 0.85, 0.95, and 1.05, shown as
green crosses.  This behavior is clearly inconsistent with the overall
behavior of the structure function extracted from electron and muon scattering.
The BCDMS data exhibit somewhat unusual behavior at large $\xi$. Above
$\xi=0.65$, the BCDMS data show little or no $Q^2$ dependence, even though one
expects noticeable QCD evolution.

\begin{figure}[htpb] 
\includegraphics[height=0.45\textwidth,angle=270]{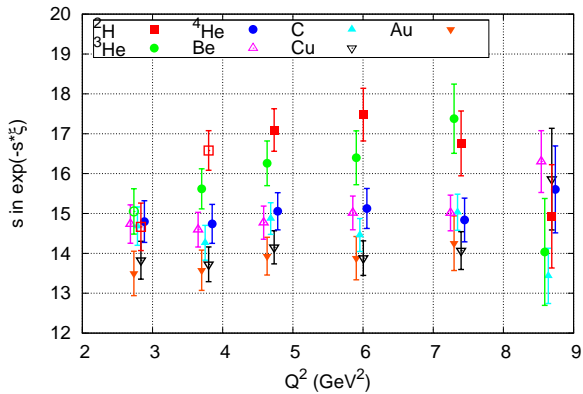}
\caption{(Color online) The slope $s$ in exp$(-s\cdot\xi)$ as a function of
$Q^2$. The targets are offset in $Q^2$ for visibility.  Open symbols
for $^2$H and $^3$He at low $Q^2$ are cases where the kinematic limit for the
nucleus ($x\approx A$) corresponds to $\xi \ltorder 1.25$.}
\label{fig:slopes_vs_q2}
\end{figure}

To quantitatively examine the falloff of our structure function at large
$\xi$, we perform a fit similar to BCDMS and CCFR. We take the
data from a fixed scattering angle, use the global fit to interpolate to a
fixed $Q^2$ value (corresponding to $\xi=1.1$), and fit
$\f20(\xi,Q^2)$ to exp$(-s\cdot\xi)$ for $1<\xi<1.25$.  The lower $\xi$ limit
is chosen to avoid regions where the quasielastic peak leads to noticeable
deviations from scaling at low $Q^2$, and the upper $\xi$ limit is chosen so
that there are data covering the full $\xi$ range for all targets and $Q^2$
values.  We take the slope extracted from the 40$\deg$ data
($Q^2$=7.35~GeV$^2$) as the main result, as this is the largest $Q^2$ value
with high statistics over the full $\xi$ range.  Data at smaller angles are
used to examine the $Q^2$ dependence of the result.

\begin{table}[htpt]
\begin{center}
\caption{Extracted values of the slopes for all nuclei.  The uncertainties
includes statistics and systematics; the latter are typically 
$\sim$0.4 and dominate the uncertainty.}
\begin{tabular}{|c|c|c|c|c|c|}
\hline
A & $Q^2=2.79$   &  3.75 &  4.68 &  5.95 &  7.35 \\				 
\hline
2  & 14.7$\pm$0.6  &  16.6$\pm$0.5 & 17.1$\pm$0.5 & 17.5$\pm$0.7&16.8$\pm$0.8\\ 
3  & 15.1$\pm$0.6  &  15.6$\pm$0.5 & 16.6$\pm$0.6 & 16.4$\pm$0.7&17.4$\pm$0.9\\ 
4  & 14.8$\pm$0.5  &  14.7$\pm$0.5 & 15.1$\pm$0.5 & 15.1$\pm$0.5&14.8$\pm$0.6\\ 
9  & 14.7$\pm$0.5  &  14.6$\pm$0.4 & 14.8$\pm$0.4 & 15.0$\pm$0.4&15.0$\pm$0.5\\ 
12  & 14.7$\pm$0.5 &  14.3$\pm$0.4 & 14.9$\pm$0.4 & 14.5$\pm$0.4&15.0$\pm$0.5\\ 
64  & 13.8$\pm$0.5 &  13.7$\pm$0.4 & 14.2$\pm$0.4 & 13.9$\pm$0.4&14.1$\pm$0.5\\ 
197 & 13.5$\pm$0.6 &  13.6$\pm$0.5 & 13.9$\pm$0.5 & 13.9$\pm$0.5&14.3$\pm$0.7\\ 
\hline
\end{tabular}
\label{tab:xi_slope_fits}
\end{center}
\end{table}

The extracted slopes are shown in Table~\ref{tab:xi_slope_fits} and
Fig.~\ref{fig:slopes_vs_q2}.  Above 4~GeV$^2$, there is no systematic $Q^2$
dependence, and at lower $Q^2$, only the $^2$H and $^3$He results change
significantly.  We observe nearly identical behavior in the high-$\xi$ falloff
for all nuclei except $^2$H and $^3$He, which have a larger slope and thus a
steeper falloff with $\xi$.

We obtain $s = 15.0 \pm 0.5$ for carbon, $s = 14.1 \pm
0.5$ for copper (our closest nucleus to the CCFR iron target), showing
that the large difference between BCDMS and CCFR is not related to the
difference in target nuclei.  Note that BCDMS and CCFR extract slopes from
$F_2(x)$ instead of $\f20(\xi)$, although the difference would increase
their slopes by less then 0.5 (0.1) for the BCDMS (CCFR).  Further
complicating direct comparison is the fact that none of these experiments
cover the same $\xi$ range.  For our new data, variations in the $\xi$ limits
of 0.05--0.1 can change the extracted slope by 0.5--1.0.

\begin{figure}[!htpb] 
\begin{center}
\includegraphics*[height=0.45\textwidth,width=7.0cm,angle=270]{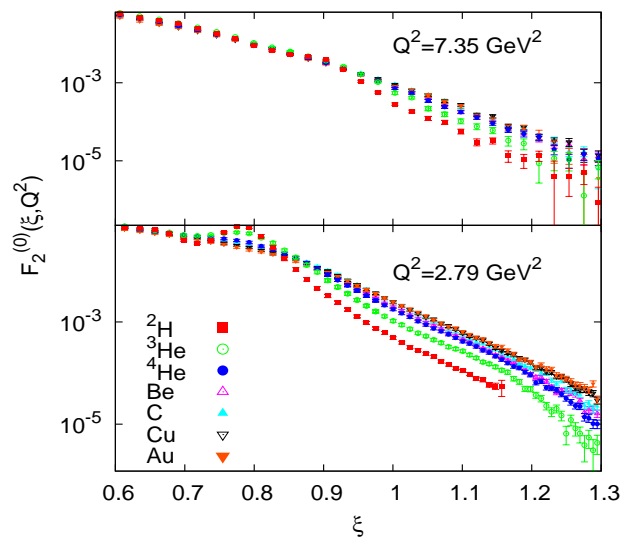}
\caption{(Color online) The extracted scaling structure function per nucleon
for all nuclei at $Q^2$=2.79 and 7.35~GeV$^2$.  The deuteron data are
kinematically limited to $x<2$, corresponding to $\xi \ltorder 1.15$ for the
low $Q^2$ setting.}
\label{fig:adep}
\end{center}
\end{figure}

We have focused on heavier nuclei for comparison to BCDMS and CCFR, but
have also obtained a significantly expanded body of data for light nuclei,
shown in Fig.~\ref{fig:adep}. At low $Q^2$, the light nuclei show a clear
quasielastic peak, while at higher $Q^2$ the peak is almost entirely washed
out.  For heavier nuclei, the extracted scaling function per nucleon is nearly
identical at all $\xi$ values.  However, for $^2$H and $^3$He, there is a
significant reduction in strength for $\xi \gtorder 1$, which is observed at
all $Q^2$ values.

In summary, we have made extensive measurements of the large $\xi$ structure
functions for $Q^2$ from 2--9~GeV$^2$.  We have extracted the scaling
structure function, $\f20(\xi,Q^2)$, and shown that it is consistent with a
nearly logarithmic $Q^2$ dependence over a significant range of $\xi$ and $Q^2$.
These new data do not show a need for extremely large contributions from
short-range correlations or other, more exotic, short range structures, as
suggested by the CCFR result.  The large $\xi$ behavior of our data is
consistent with the BCDMS results, but our results extend to significantly
higher $\xi$ values, where one expects to be most sensitive to short range
structure~\cite{frankfurt88,sargsian03}. The new data, covering a range of
nuclei at large $\xi$, can be used to directly constrain calculations of the
effect of short-range correlations or multi-quark configurations in our
kinematic regime for a wide range of nuclei.  A future 12 GeV JLab
measurement~\cite{e1206015} will double the $Q^2$ range for these $\xi$ values,
moving to a region where we can directly extract the parton distributions of
the super-fast quarks in nuclei.

\begin{acknowledgments}

We thank the JLab technical staff and accelerator
division for their contributions.  This work supported in part by the NSF
and DOE, including grant NSF-0244899 and DOE contracts DE-FG02-96ER40950,
DE-AC02-06CH11357 and DE-AC05-06OR23177 under which JSA, LLC operates JLab,
and the South African NRF.

\end{acknowledgments}

\bibliography{prl_highx}

\begin{thebibliography}{21}
\expandafter\ifx\csname natexlab\endcsname\relax\def\natexlab#1{#1}\fi
\expandafter\ifx\csname bibnamefont\endcsname\relax
  \def\bibnamefont#1{#1}\fi
\expandafter\ifx\csname bibfnamefont\endcsname\relax
  \def\bibfnamefont#1{#1}\fi
\expandafter\ifx\csname citenamefont\endcsname\relax
  \def\citenamefont#1{#1}\fi
\expandafter\ifx\csname url\endcsname\relax
  \def\url#1{\texttt{#1}}\fi
\expandafter\ifx\csname urlprefix\endcsname\relax\def\urlprefix{URL }\fi
\providecommand{\bibinfo}[2]{#2}
\providecommand{\eprint}[2][]{\url{#2}}

\bibitem[{\citenamefont{Schienbein et~al.}(2008)}]{schienbein08}
\bibinfo{author}{\bibfnamefont{I.}~\bibnamefont{Schienbein}}
  \bibnamefont{et~al.}, \bibinfo{journal}{J. Phys.}
  \textbf{\bibinfo{volume}{G35}}, \bibinfo{pages}{053101}
  (\bibinfo{year}{2008}).

\bibitem[{\citenamefont{Geesaman et~al.}(1995)\citenamefont{Geesaman, Saito,
  and Thomas}}]{geesaman95}
\bibinfo{author}{\bibfnamefont{D.~F.} \bibnamefont{Geesaman}},
  \bibinfo{author}{\bibfnamefont{K.}~\bibnamefont{Saito}}, \bibnamefont{and}
  \bibinfo{author}{\bibfnamefont{A.~W.} \bibnamefont{Thomas}},
  \bibinfo{journal}{Ann. Rev. Nucl. Sci.} \textbf{\bibinfo{volume}{45}},
  \bibinfo{pages}{337} (\bibinfo{year}{1995}).

\bibitem[{\citenamefont{Frankfurt et~al.}(1993)\citenamefont{Frankfurt,
  Strikman, Day, and Sargsian}}]{frankfurt93}
\bibinfo{author}{\bibfnamefont{L.~L.} \bibnamefont{Frankfurt}},
  \bibinfo{author}{\bibfnamefont{M.~I.} \bibnamefont{Strikman}},
  \bibinfo{author}{\bibfnamefont{D.~B.} \bibnamefont{Day}}, \bibnamefont{and}
  \bibinfo{author}{\bibfnamefont{M.}~\bibnamefont{Sargsian}},
  \bibinfo{journal}{Phys. Rev. C} \textbf{\bibinfo{volume}{48}},
  \bibinfo{pages}{2451} (\bibinfo{year}{1993}).

\bibitem[{\citenamefont{Sargsian et~al.}(2003)}]{sargsian03}
\bibinfo{author}{\bibfnamefont{M.~M.} \bibnamefont{Sargsian}}
  \bibnamefont{et~al.}, \bibinfo{journal}{J. Phys.}
  \textbf{\bibinfo{volume}{G29}}, \bibinfo{pages}{R1} (\bibinfo{year}{2003}).

\bibitem[{\citenamefont{Benvenuti et~al.}(1994)}]{benvenuti94}
\bibinfo{author}{\bibfnamefont{A.}~\bibnamefont{Benvenuti}}
  \bibnamefont{et~al.} (\bibinfo{collaboration}{BCDMS}), \bibinfo{journal}{Z.
  Phys.} \textbf{\bibinfo{volume}{C63}}, \bibinfo{pages}{29}
  (\bibinfo{year}{1994}).

\bibitem[{\citenamefont{Vakili et~al.}(2000)}]{vakili00}
\bibinfo{author}{\bibfnamefont{M.}~\bibnamefont{Vakili}} \bibnamefont{et~al.}
  (\bibinfo{collaboration}{CCFR}), \bibinfo{journal}{Phys. Rev. D}
  \textbf{\bibinfo{volume}{61}}, \bibinfo{pages}{052003}
  (\bibinfo{year}{2000}).

\bibitem[{\citenamefont{Rozynek and Birse}(1988)}]{rozynek88}
\bibinfo{author}{\bibfnamefont{J.}~\bibnamefont{Rozynek}} \bibnamefont{and}
  \bibinfo{author}{\bibfnamefont{M.~C.} \bibnamefont{Birse}},
  \bibinfo{journal}{Phys. Rev.} \textbf{\bibinfo{volume}{C38}},
  \bibinfo{pages}{2201} (\bibinfo{year}{1988}).

\bibitem[{\citenamefont{Frankfurt and Strikman}(1988)}]{frankfurt88}
\bibinfo{author}{\bibfnamefont{L.~L.} \bibnamefont{Frankfurt}}
  \bibnamefont{and} \bibinfo{author}{\bibfnamefont{M.~I.}
  \bibnamefont{Strikman}}, \bibinfo{journal}{Phys. Rept.}
  \textbf{\bibinfo{volume}{160}}, \bibinfo{pages}{235} (\bibinfo{year}{1988}).

\bibitem[{\citenamefont{Bosted et~al.}(1992)}]{bosted92}
\bibinfo{author}{\bibfnamefont{P.}~\bibnamefont{Bosted}} \bibnamefont{et~al.},
  \bibinfo{journal}{Phys. Rev. C} \textbf{\bibinfo{volume}{46}},
  \bibinfo{pages}{2505} (\bibinfo{year}{1992}).

\bibitem[{\citenamefont{Arrington et~al.}(1996)}]{arrington96}
\bibinfo{author}{\bibfnamefont{J.}~\bibnamefont{Arrington}}
  \bibnamefont{et~al.}, \bibinfo{journal}{Phys. Rev. C}
  \textbf{\bibinfo{volume}{53}}, \bibinfo{pages}{2248} (\bibinfo{year}{1996}).

\bibitem[{\citenamefont{Day et~al.}(1979)}]{day79}
\bibinfo{author}{\bibfnamefont{D.~B.} \bibnamefont{Day}} \bibnamefont{et~al.},
  \bibinfo{journal}{Phys. Rev. Lett.} \textbf{\bibinfo{volume}{43}},
  \bibinfo{pages}{1143} (\bibinfo{year}{1979}).

\bibitem[{\citenamefont{Filippone et~al.}(1992)}]{filippone92}
\bibinfo{author}{\bibfnamefont{B.~W.} \bibnamefont{Filippone}}
  \bibnamefont{et~al.}, \bibinfo{journal}{Phys. Rev. C}
  \textbf{\bibinfo{volume}{45}}, \bibinfo{pages}{1582} (\bibinfo{year}{1992}).

\bibitem[{\citenamefont{Arrington et~al.}(2001)}]{arrington01}
\bibinfo{author}{\bibfnamefont{J.}~\bibnamefont{Arrington}}
  \bibnamefont{et~al.}, \bibinfo{journal}{Phys. Rev. C}
  \textbf{\bibinfo{volume}{64}}, \bibinfo{pages}{014602}
  (\bibinfo{year}{2001}).

\bibitem[{\citenamefont{Seely et~al.}(2009)}]{seely09}
\bibinfo{author}{\bibfnamefont{J.}~\bibnamefont{Seely}} \bibnamefont{et~al.},
  \bibinfo{journal}{Phys. Rev. Lett.} \textbf{\bibinfo{volume}{103}},
  \bibinfo{pages}{202301} (\bibinfo{year}{2009}).

\bibitem[{\citenamefont{Fomin}(2007)}]{fominphd}
\bibinfo{author}{\bibfnamefont{N.}~\bibnamefont{Fomin}}, Ph.D. thesis,
  \bibinfo{school}{University of Virginia} (\bibinfo{year}{2007}),
  \bibinfo{note}{arXiv:0812.2144}.

\bibitem[{\citenamefont{Nachtmann}(1973)}]{nachtmann73}
\bibinfo{author}{\bibfnamefont{O.}~\bibnamefont{Nachtmann}},
  \bibinfo{journal}{Nucl. Phys.} \textbf{\bibinfo{volume}{B63}},
  \bibinfo{pages}{237} (\bibinfo{year}{1973}).

\bibitem[{\citenamefont{Arrington
  et~al.}(2006{\natexlab{a}})\citenamefont{Arrington, Ent, Keppel, Mammei, and
  Niculescu}}]{arrington06a}
\bibinfo{author}{\bibfnamefont{J.}~\bibnamefont{Arrington}},
  \bibinfo{author}{\bibfnamefont{R.}~\bibnamefont{Ent}},
  \bibinfo{author}{\bibfnamefont{C.~E.} \bibnamefont{Keppel}},
  \bibinfo{author}{\bibfnamefont{J.}~\bibnamefont{Mammei}}, \bibnamefont{and}
  \bibinfo{author}{\bibfnamefont{I.}~\bibnamefont{Niculescu}},
  \bibinfo{journal}{Phys. Rev. C} \textbf{\bibinfo{volume}{73}},
  \bibinfo{pages}{035205} (\bibinfo{year}{2006}{\natexlab{a}}).

\bibitem[{\citenamefont{Melnitchouk et~al.}(2005)\citenamefont{Melnitchouk,
  Ent, and Keppel}}]{melnitchouk05}
\bibinfo{author}{\bibfnamefont{W.}~\bibnamefont{Melnitchouk}},
  \bibinfo{author}{\bibfnamefont{R.}~\bibnamefont{Ent}}, \bibnamefont{and}
  \bibinfo{author}{\bibfnamefont{C.}~\bibnamefont{Keppel}},
  \bibinfo{journal}{Phys. Rept.} \textbf{\bibinfo{volume}{406}},
  \bibinfo{pages}{127} (\bibinfo{year}{2005}).

\bibitem[{\citenamefont{Whitlow et~al.}(1992)\citenamefont{Whitlow, Riordan,
  Dasu, Rock, and Bodek}}]{whitlow92}
\bibinfo{author}{\bibfnamefont{L.~W.} \bibnamefont{Whitlow}},
  \bibinfo{author}{\bibfnamefont{E.~M.} \bibnamefont{Riordan}},
  \bibinfo{author}{\bibfnamefont{S.}~\bibnamefont{Dasu}},
  \bibinfo{author}{\bibfnamefont{S.}~\bibnamefont{Rock}}, \bibnamefont{and}
  \bibinfo{author}{\bibfnamefont{A.}~\bibnamefont{Bodek}},
  \bibinfo{journal}{Phys. Lett.} \textbf{\bibinfo{volume}{B282}},
  \bibinfo{pages}{475} (\bibinfo{year}{1992}).

\bibitem[{\citenamefont{Gomez et~al.}(1994)}]{gomez94}
\bibinfo{author}{\bibfnamefont{J.}~\bibnamefont{Gomez}} \bibnamefont{et~al.},
  \bibinfo{journal}{Phys. Rev. D} \textbf{\bibinfo{volume}{49}},
  \bibinfo{pages}{4348} (\bibinfo{year}{1994}).

\bibitem[{\citenamefont{Arrington
  et~al.}(2006{\natexlab{b}})\citenamefont{Arrington, Day et~al.}}]{e1206015}
\bibinfo{author}{\bibfnamefont{J.}~\bibnamefont{Arrington}},
  \bibinfo{author}{\bibfnamefont{D.~B.} \bibnamefont{Day}},
  \bibnamefont{et~al.} (\bibinfo{year}{2006}{\natexlab{b}}),
  \bibinfo{note}{{J}Lab Proposal E12-06-105}.

\end{thebibliography}

\end{document}